\documentclass[aps,showpacs,floatfix,nofootinbib,11pt]{revtex4-1}

\usepackage[]{graphicx}
\usepackage{epstopdf}
\usepackage[intlimits]{amsmath}
\usepackage{upgreek}
\usepackage[utf8]{inputenc}
\usepackage[colorlinks,linkcolor=blue,citecolor=blue,urlcolor=blue,breaklinks=true]{hyperref}
\usepackage{amssymb,soul,physics,enumitem,cleveref}
\usepackage[table]{xcolor}

\newcommand{\rv}{\ensuremath{\vb{x}}}
\newcommand{\CoM}{\ensuremath{_\text{CoM}}}
\newcommand{\tinit}{\ensuremath{t_\text{i}}}
\newcommand{\tfinal}{\ensuremath{t_\text{f}}}
\newcommand{\epsNot}{\ensuremath{\varepsilon_0}}
\newcommand{\Coul}{\ensuremath{_\text{C}}}
\newcommand{\Poin}{\ensuremath{_\text{P}}}
\newcommand{\PZW}{\ensuremath{_\text{\tiny PZW}}}
\newcommand{\trans}{\ensuremath{_\bot}}
\newcommand{\longi}{\ensuremath{_\Vert}}
\newcommand{\Lcal}{\ensuremath{\mathcal{L}}}
\newcommand{\bs}[1]{\boldsymbol{#1}}
\newcommand{\timeint}{\int_{\tinit}^{\tfinal} \dd{t}}
\newcommand{\spaceint}{\int \dd[3]{x}}
\newcommand{\sint}{\int_{0}^{1} \dd{s}}

\definecolor{LightPink}{rgb}{1,.9,.9}

\begin{document}

\title{The gauge-invariant Lagrangian, the Power-Zienau-Woolley picture, and the choices of field momenta in nonrelativistic quantum electrodynamics}
\author{A. Vukics}
\email{vukics.andras@wigner.mta.hu}
\author{G. Kónya}
\author{P. Domokos}
\affiliation{Wigner Research Centre for Physics, H-1525 Budapest, P.O. Box 49., Hungary}

\date{\today}

\begin{abstract}
We show that the Power-Zienau-Woolley picture of the electrodynamics of nonrelativistic neutral particles (atoms) can be derived from a gauge-invariant Lagrangian without making reference to any gauge whatsoever in the process. This equivalence is independent of choices of canonical field momentum or quantization strategies. In the process, we emphasize that in nonrelativistic (quantum) electrodynamics, the all-time appropriate generalized coordinate for the field is the transverse part of the vector potential, which is itself gauge invariant, and the use of which we recommend regardless of the choice of gauge, since in this way it is possible to sidestep most issues of constraints. Furthermore, we point out a freedom of choice for the conjugate momenta in the respective pictures, the conventional choices being good ones in the sense that they drastically reduce the set of system constraints.
\end{abstract}

\maketitle

\tableofcontents

\section{Introduction}


Let us consider a neutral atom consisting of nonrelativistic point particles coupled to the electromagnetic field. To describe this system either classically or quantum mechanically, one can use different formulations of the same dynamics. One approach uses the minimal-coupling Hamiltonian, while another approach uses the multipolar Hamiltonian. This latter Hamiltonian is conventionally derived through a unitary transformation, which is referred to as the Power-Zienau-Woolley (PZW) transformation.

In a recent article, Rousseau and Felbacq \cite{Rousseau2017} argue that the derivation of the multipolar Hamiltonian based on the PZW transformation is inconsistent, based on consideration of gauges, constraints, and choices of canonical momenta. These claims call into question a significant body of work in nonrelativistic quantum electrodynamics, research including a recent paper of ours \cite{Power1959427,Woolley1971,Woolley1974488,Woolley19802795,Babiker1983,Power19832649,Ackerhalt1984116,Vukics2014}, as well as textbooks \cite{CDG,Milonni1994,Compagno1995}. In this paper, we clearly refute these concerns.

In particular, we show (\cref{sec:pzwOnTheAction}) that the PZW transformation can be directly performed at the level of the Lagrangian formalism, even before any Hamiltonians are constructed. Furthermore, it is possible to use a gauge-invariant form of the Lagrangian of the minimal-coupling picture as the starting point of such a derivation. This shows that the PZW transformation cannot be declared inconsistent based on arguments about gauge or canonical momentum, since the equivalence of the PZW picture with the minimal-coupling one is demonstrated in a completely gauge-invariant manner at the level of the Lagrangian, even before the canonical momentum variables and the Hamiltonian are introduced (which we will do only in \cref{sec:pzwHamiltonian}).

Our approach here is based on the use of the \emph{transverse part of the vector potential} (\(\vb A\trans\)) as generalized field coordinate, which is also gauge invariant (a gauge transformation changes only the scalar potential and the longitudinal part of the vector potential). We advocate the use of this coordinate in whatever picture or gauge. This choice is superior to using the full vector potential (\(\vb A\)) in some gauge not only because the gauge-invariance of the treatment would be lost, but also because with \(\vb A\) as coordinate, the particle and field degrees of freedom would be mixed, leading to an inconvenient set of system constraints (or, nontrivial commutation relations in the quantum case). With our choice, the treatment of constraints is not an issue, and there remains only a single nontrivial commutation relation, which is moreover the same in both the minimal-coupling and PZW pictures.

In the case of constrained systems, the real power of the Lagrangian formalism is manifested when the constraints can be treated by an appropriate choice of reduced coordinates that bijectively cover the hyperplane in configuration space (c-space) defined by the constraints (this hyperplane is called configuration manifold – c-manifold).\footnote{When this is not possible, the more general method of the Lagrangian multiplier must be used.} Then, one does not need to worry about the constraints anymore, since they are encoded in the very choice of generalized coordinates. This is the case with the ideal pendulum, where we can choose the angle of the pendulum instead of the Cartesian coordinates to cover the c-manifold. This is also the case in the eletrodynamics of nonrelativistic charged particles, where we can choose \(\vb A\trans\) plus the particle coordinates to cover the c-manifold in a bijective manner.\footnote{This is not the case in the electrodynamics of relativistic particles (fields), where the covariant formulation necessitates the use of the full potential four-vector.}

\begin{figure}
 \includegraphics[width=\linewidth]{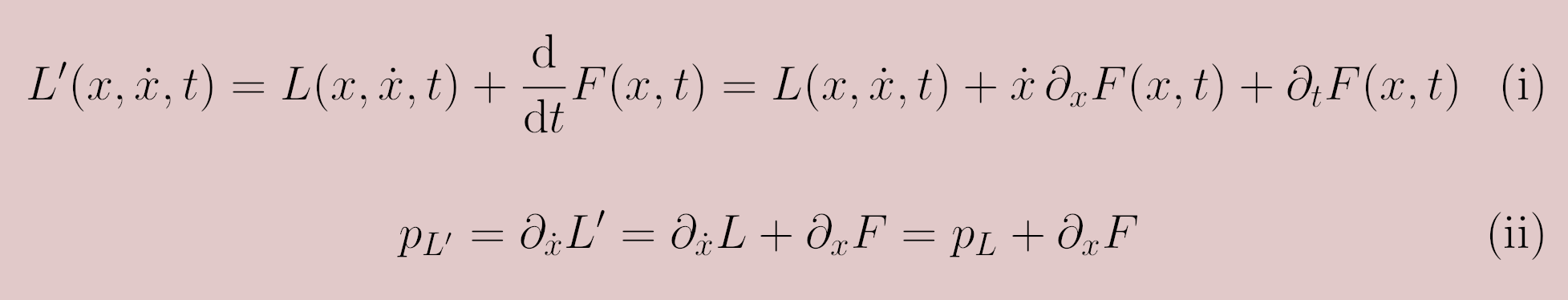}
 \caption{Change of conjugate momentum under an equivalent transformation of the Lagrangian \(L\) defined by \(F\), a function of the generalized coordinates and time}
 \label{fig:generalMomentumTrafo}
\end{figure}

In this paper, the PZW transformation is introduced as an equivalent transformation of the Lagrangian. Such transformations do not change the generalized coordinates because these are in fact part of the definition of the Lagrangian problem. Without saying what the generalized coordinates and velocities are, as a function(al) of which the Lagrangian is considered, the problem is ill-defined. On the other hand, the momenta conjugate to the coordinates do change under an equivalent transformation, which is the case already in point mechanics, cf. \cref{fig:generalMomentumTrafo}.\footnote{In the original formulation, the PZW transformation was treated as a unitary transformation in the quantum case. In this case, the transformation operator commutes with \(\vb A\trans\) so that the coordinate remains unchanged in this formulation as well.}

Another important observation is that once the set of generalized coordinates is fixed, the conjugate field momenta still exhibit a certain ambiguity, which is unknown in point mechanics, being field-theoretic in nature. \Cref{sec:momentumAmbiguity} is devoted to the analysis of this phenomenon. The point is that without changing the Lagrangian (and hence without changing the gauge), due to that spatial integrals of certain combinations of fields satisfying certain relations (including gauge conditions) vanish, different functional derivatives can be extracted. If one works in the minimal coupling picture, a longitudinal vector field can be freely added to the canonical momentum. This allows one to replace the electric field with its transverse part. In the PZW picture, an analogous freedom is present: we find that here the displacement field and the transverse electric field are two legitimate choices for the canonical field momentum.

\section{Fundamentals}

We consider a neutral atom consisting of nonrelativistic point particles and interacting with the electromagnetic field.\footnote{The reason why it is important to consider only one atom will be explained in \cref{fn:oneatom}.} Although the present section largely consists of textbook material, we give a succint account of these foundations that will be pertinent to the delicate questions of canonical coordinates-momenta and gauges.

\subsection{Maxwell equations}

The two homogeneous Maxwell equations read:
\begin{subequations}
\begin{align}
\curl{\vb{E}} &= - \partial_t \vb{B} \; ,\\
\div{\vb{B}} &= 0 \; ,
\end{align}
and the two inhomogeneous ones:
\begin{align}
\curl{\vb{B}} &= \mu_0 \, \vb{j} + \epsNot \, \mu_0 \, \partial_t \vb{E} \; ,\\
\label{eq:MaxwellTwo}
\div{\vb{E}} &= \frac{1}{\epsNot} \, \rho \; .
\end{align}
\end{subequations}
The source terms of the equations are the charge and current densities, which for point-like particles read respectively:
\begin{subequations}
\begin{align}
\label{eq:chargeDensity}
\rho (\rv,t) &= \sum_{\alpha=0}^{Z} \, q_{\alpha} \, \delta \qty( \rv - \rv_\alpha (t)) \,   \; ,\\
\vb{j} (\rv,t) &= \sum_{\alpha=0}^{Z} \, q_{\alpha} \, \dot{\rv}_\alpha (t) \, \delta \qty( \rv - \rv_\alpha (t) ) \; .
\end{align}
\end{subequations}
Here \(\alpha\) indexes the particles constituting the atom of atomic number \(Z\), with \(\alpha=0\) denoting the nucleus. From the two inhomogeneous Maxwell equations, it can be shown that they satisfy the continuity equation:
\begin{equation}
\label{eq:continuity}
\partial_t \rho + \div{\vb{j}} = 0 \; ,
\end{equation}
which is needed also for the consistency of the Maxwell equations. The particle motion is governed by the Lorentz force:
\begin{equation}
m_{\alpha} \, \ddot{\rv}_\alpha (t) = q_{\alpha} \big[ \vb{E} (\rv_\alpha(t),t) + \dot{\rv}_\alpha (t) \times \vb{B} (\rv_\alpha(t),t) \big] \; .
\end{equation} 

\subsection{The potentials and the Coulomb gauge}

The two homogeneous Maxwell equations can be automatically satisfied by deriving the fields from potentials as
\begin{subequations}
\label{eq:fieldsFromPotentials}
\begin{align}
\label{eq:fieldsFromPotentialsOne}
\vb{E} &= -\partial_t \vb{A} - \grad{\Phi} \; ,\\
\vb{B} &= \curl{\vb{A}} \; ,
\end{align}
\end{subequations}
\(\Phi\) being the scalar, while \(\vb{A}\) the vector potential. They can be subjected to gauge transformations that leave the physical fields unchanged:
\begin{subequations}
\label{eq:gaugeTrafo}
\begin{align}
\label{eq:gaugeTrafoPhi}
\Phi' &= \Phi + \partial_t \chi \; ,\\
\label{eq:gaugeTrafoA}
\vb{A}' &= \vb{A} - \grad{\chi} \; . 
\end{align}
\end{subequations}
Here, \(\chi\) is an arbitrary scalar field. Such transformations constitute the main theme of the discussion at hand.

The most common choice of gauge in the electrodynamics of atoms is the Coulomb gauge, defined by
\begin{equation}
\div{\vb{A}\Coul} =0  \; .
\end{equation}
Then, the transverse and longitudinal components of the electric field read:
\begin{subequations}
\begin{align}
\label{eq:dtA}
\vb{E}\trans &= - \partial_t \, \vb{A}\Coul \; ,\\
\label{eq:electrostatic1}
\vb{E}\longi &= - \grad{\Phi\Coul} \; .
\end{align}
\end{subequations}
Where by definition the transverse component is divergence- while the longitudinal is curl-free (Helmholtz decomposition). Moreover, according to the Maxwell equation \labelcref{eq:MaxwellTwo} we have 
\begin{equation}
\label{eq:electrostatic2}
\laplacian{\Phi\Coul} = -\frac{1}{\epsNot} \, \rho \quad\Longrightarrow\quad \Phi\Coul\qty(\vb x,t)=\frac1{4\pi\epsNot}\int \dd[3]{x'}\frac{\rho\qty(\vb x',t)}{\abs{\vb x'-\vb x}}=
\frac1{4\pi\epsNot}\sum_{\alpha=0}^Z\frac{q_\alpha}{\abs{\vb x_\alpha(t)-\vb x}},
\end{equation}
meaning that in Coulomb gauge, the scalar potential is not a dynamical variable, but it is fixed to the charge density.

Let us write the solution of the electrostatic problem in a gauge-invariant form as:
\begin{equation}
 \label{eq:electrostaticGaugeInvariant}
 \vb E\longi(\rv,t)=-\frac1{4\pi\epsNot}\sum_{\alpha=0}^Zq_\alpha\frac{\qty(\vb x_\alpha(t)-\vb x)}{\abs{\vb x_\alpha(t)-\vb x}^3}\;.
\end{equation}

\subsection{The Lagrangian}
The Lagrangian of nonrelativistic electrodynamics is usually written in the form
\begin{equation}
\label{eq:fundamentalAction}
L(\vb x_\alpha,\dot{\vb x}_\alpha,?) = \sum_{\alpha=0}^{Z} \frac{m_{\alpha}}{2} \, \dot{\rv}_\alpha^2 (t)
+ \spaceint \qty[ \frac{\epsNot}{2} \, \vb{E}^2  - \frac{1}{2 \, \mu_0} \, \vb{B}^2  ] 
+ \spaceint\, \qty\big[ \vb{j}  \cdot \vb{A}  - \rho  \, \Phi  ] \; .
\end{equation}
Although Maxwell’s equations and the Lorentz force can be derived \emph{formally} by the variational method hence, the Lagrangian does not make sense \emph{per se,} only as function (functional) of well-defined generalized coordinates. In this form, if the field dynamics was treated with the full potentials as coordinates then there would be a redundancy and interdependence of coordinates (cf. \cite{CDG} Section II.B.3.c). Indeed, the Maxwell equations exhibit only four dynamical field variables (the two components each of the transverse part of the two physical fields), while the \emph{a priori} description by potentials gives eight (three components of the vector and one of the scalar potential plus as many components of generalized velocities).

An appropriate way to reduce the redundancy is to fix the gauge to the Coulomb gauge:
\begin{multline}
\label{eq:CoulombAction}
L\Coul=L(\vb x_\alpha,\dot{\vb x}_\alpha,\vb A\Coul,\partial_t\vb A\Coul)
=\sum_{\alpha=0}^{Z} \qty(\frac{m_{\alpha}}{2} \, \dot{\rv}_\alpha^2(t) - \frac1{4\pi\epsNot}\sum_{\beta=\alpha+1}^Z \frac{q_\alpha\,q_\beta}{\abs{\vb x_\alpha-\vb x_\beta}} )
\\+ \spaceint \qty[ \frac{\epsNot}{2} \, \vb{E}\trans^2  - \frac{1}{2 \, \mu_0} \, \vb{B}^2  ]
+ \spaceint \,\vb{j}\trans  \cdot \vb{A}\Coul.
\end{multline}
Here, to calculate the electrostatic term (second term within the summation over \(\alpha\)), the terms containing \(\vb E\longi^2\) and \(\rho\,\Phi\) had to be rewritten with the help of \cref{eq:electrostatic1,eq:electrostatic2}. In this form of the Lagrangian, the first term defines the atom, the second the field, and the third the interaction between these two. All these terms are gauge invariant; furthermore, the generalized field coordinate is also gauge invariant in the sense that \(\vb A\Coul=\vb A\trans\) and \(\vb A\trans\) is gauge invariant because the transformation \labelcref{eq:gaugeTrafoA} touches only the longitudinal part of the vector potential.

Therefore, although we invoked the Coulomb gauge as a methodological step to obtain \cref{eq:CoulombAction} from \cref{eq:fundamentalAction}, there is no need to make any consideration of gauge when working with this form of the Lagrangian. In fact, we could have solved the electrostatic part of the problem in any gauge, and arrive at the same Lagrangian. The reason why the Coulomb gauge was invoked is solely because it is in this gauge that the electrostatic problem is the easiest.

To emphasize the fact of term-by-term gauge invariance, the use of gauge-invariant coordinate, and the overall irrelevance of the choice of gauge, we reexpress the Lagrangian \labelcref{eq:CoulombAction} in the manifestly gauge invariant form:
\begin{multline}
\label{eq:ultimateAction}
L(\vb x_\alpha,\dot{\vb x}_\alpha,\vb A\trans,\partial_t\vb A\trans)
=\sum_{\alpha=0}^{Z}\qty( \frac{m_{\alpha}}{2} \, \dot{\rv}_\alpha^2(t) - \frac1{4\pi\epsNot}\sum_{\beta=\alpha+1}^Z \frac{q_\alpha\,q_\beta}{\abs{\vb x_\alpha-\vb x_\beta}} )\\
+ \spaceint \qty[ \frac{\epsNot}{2} \, \vb{E}\trans^2  - \frac{1}{2 \, \mu_0} \, \vb B^2  ]
+ \spaceint \,\vb{j}\trans  \cdot \vb{A}\trans,
\end{multline}
where the field coordinate \(\vb A\trans\) denotes the transverse part of the vector potential.\footnote{It is worth noting because it sometimes leads to misundestandings that there is no general formula that expresses gauge transformation on the Lagrangian level. Indeed, if we consider the Lagrangian \labelcref{eq:fundamentalAction}, then we see that its transformation under a gauge transformation is given as 
\begin{equation}
\label{eq:mertektrafo_hatason}
\Delta L = - \spaceint \,\qty\big[ \vb{j}  \cdot \grad{\chi } + \rho \, \partial_t \chi ]
= \spaceint\, \qty\big[ \div{\vb{j} } + \partial_t \rho]
\chi  - \dv{t}\spaceint \rho  \, \chi = -\dv{t}\sum_{\alpha=0}^{Z} q_\alpha \chi (\rv_\alpha(t),t)\; ,
\end{equation}
where we have performed integration by parts both in space and time. After the second equality sign, the first correction term vanishes due to the continuity \cref{eq:continuity}, while the second can be evaluated through \cref{eq:chargeDensity} to obtain the final expression. So this is a nonzero change, which is however a special case of equivalent transformations, cf. \cref{eq:equivalentActionTrafo}. On the other hand, the change of the Lagrangian \labelcref{eq:ultimateAction} under gauge transformation is zero.}

The Lagrangian density \(\Lcal\) for the field can be introduced through the definition:
\begin{equation}
 L=L_\text{matter} + \spaceint\, \qty(\Lcal_\text{field} + \Lcal_\text{interaction}).
\end{equation}

\section{The Power-Zienau-Woolley picture}

Originally, the Power-Zienau-Woolley transformation was introduced as a unitary transformation from the minimal-coupling Hamiltonian into the so-called multipolar Hamiltonian. Since this transformation acts on the Hamiltonian and other operators expressing physical quantities, as well as the state vectors forming the Hilbert space, we refer to the resulting description as the Power-Zienau-Woolley \emph{picture}.

In this section, starting from the gauge-invariant Lagrangian \labelcref{eq:ultimateAction}, we present the PZW transformation as an equivalent transformation of the Lagrangian under the form (cf. \cite{landau} Section I.1):
\begin{subequations}
\label{eq:equivalentActionTrafo}
\begin{equation}
 L'=L+\dv{t}\qty(\parbox{14.75ex}{\scriptsize anything that depends only on time and the generalized coordinates})
\end{equation}
In the language of the action, such an equivalent transformation reads
\begin{equation}
 S' = \timeint L' = S + \qty(\parbox{18ex}{\scriptsize anything that depends only on \tinit\ and \tfinal\ and the values of coordinates at these instants}).
\end{equation}
\end{subequations}
The action \(S'\) is equivalent to \(S\) in the sense that it leads to the same equations of motion, due to the fact that variations of generalized coordinates and velocities vanish at the extremal time instants \(\tinit\) and \(\tfinal\) according to the variational lore. 

First, we introduce the polarization and magnetization fields which are key quantities in the PZW picture to express the charge and current densities.

\subsection{Polarization and magnetization fields}

Let us start with the charge density. Upon introducing the polarization field as
\begin{equation}
\label{eq:polarizationField}
\vb{P} (\rv,t) = \sum_{\alpha=0}^{Z} q_{\alpha} \, \vb*{\xi}_\alpha (t) \, \sint \, \delta \qty( \rv - \rv\CoM - s \, \vb*{\xi}_\alpha (t) ) \; ,
\end{equation}
where \(\rv\CoM\) is the position of the atomic center-of-mass and \(\vb*{\xi}_\alpha\)s are the relative coordinates, it is found that
\begin{equation}
\label{eq:chargesFromPolarization}
\rho = - \div{\vb{P}} \; .
\end{equation}
For the details of the derivation in a distribution-theoretic approach, cf. \Cref{app:distros}. For the physical picture behind the polarization field, cf. \cite{CDG} Section IV.C.1.

In the following, we assume that the atomic nucleus is so heavy that it sits at the center of mass – which we equate with the origin – and is immobile. This is merely in order to simplify notation, e.g.:
\begin{equation}
\label{eq:polarization}
\vb{P} (\rv,t) = \sum_{\alpha=1}^{Z} q_{\alpha} \, \rv_\alpha (t) \, \sint \, \delta \qty( \rv - s \, \rv_\alpha (t)  ) \; ,
\end{equation}

In the next step, we play a similar game with the current density. Introducing the magnetization field as:
\begin{equation}
\label{eq:magnetization}
\vb{M} (\rv,t) = \sum_{\alpha=1}^{Z} q_{\alpha} \; \rv_\alpha (t) \times \dot{\rv}_\alpha (t) \,
\sint \, s \, \delta \qty( \rv - s \, \rv_\alpha (t)  ) \; ,
\end{equation}
we find that
\begin{equation}
\label{eq:currentDensityFromMagnetization}
\vb{j} = \partial_t \vb{P} + \curl{\vb{M}} \; .
\end{equation}
That is, the current density consists of two terms, one related to the electric polarization, the other to the magnetization of the atom.


\subsection{Power-Zienau-Woolley transformation on the Lagrangian}
\label{sec:pzwOnTheAction}
The electrostatic term in the Lagrangian \labelcref{eq:ultimateAction} can be rewritten with the help of \(\vb E\longi=-\epsNot\vb P\longi\) (which follows from \cref{eq:MaxwellTwo,eq:chargesFromPolarization}) to give
\begin{equation}
\frac1{4\pi\epsNot}\sum_{\underset{\alpha<\beta\leq Z}{\alpha=0}}^Z\frac{q_\alpha\,q_\beta}{\abs{\vb x_\alpha-\vb x_\beta}}=\spaceint \frac{1}{2} \, \rho  \, \Phi\Coul  = - \spaceint \, \frac{\epsNot}{2} \, \vb{E}\longi^2  - \spaceint \, \vb{P}\longi \cdot \vb{E}\longi \; .
\end{equation}
Then, using \cref{eq:currentDensityFromMagnetization}, we can rewrite \cref{eq:ultimateAction} as
\begin{multline}
L = \sum_{\alpha=0}^{Z} \frac{m_{\alpha}}{2} \,  \dot{\rv}_\alpha^2 (t)
+ \spaceint\,\qty[ \frac{\epsNot}{2} \, \vb{E}^2  - \frac{1}{2 \, \mu_0} \, \vb{B}^2   ] \\
+ \spaceint\, \qty\bigg{ \partial_t \vb{P}\trans  \cdot \vb{A}\trans
+ \qty\big[\curl{\vb{M}} ] \cdot \vb{A}\trans}
+ \spaceint \, \vb{P}\longi  \cdot \vb{E}\longi  \; .
\end{multline}

Now let us perform the PZW transformation in the form of an equivalent Lagrangian transformation \labelcref{eq:equivalentActionTrafo}:
\begin{multline}
\label{eq:PZW_trafo}
L\PZW = L - \dv{t}\spaceint \,\vb{P}\trans  \cdot \vb{A}\trans\\=L - \spaceint \qty(\partial_t\vb{P}\trans)  \cdot \vb{A}\trans-\spaceint \vb{P}\trans\cdot\qty(\partial_t\vb{A}\trans).
\end{multline}
Perform the following integration by part \(\spaceint \, \qty\big[\curl{\vb{M}} ]\cdot \vb{A}\trans = \spaceint \, \vb{M}  \cdot \qty\big[\curl{\vb{A}\trans}]\) and substituting the physical fields in place of the vector potential, we finally obtain the new Lagrangian:
\begin{equation}
\label{eq:PZW_hatas}
L\PZW = \sum_{\alpha=0}^{Z} \, \frac{m_{\alpha}}{2} \, \dot{\rv}_\alpha^2 (t) 
+ \spaceint\, \qty[ \frac{\epsNot}{2} \, \vb{E}^2  - \frac{1}{2 \, \mu_0} \, \vb{B}^2  ]
+ \spaceint \,\qty\big[ \vb{P}  \cdot \vb{E}  + \vb{M}  \cdot \vb{B}  ] \; .
\end{equation}

For completeness, in the following subsection we derive the familiar PZW Hamiltonian, thereby proving \emph{a posteriori} that this is indeed the Lagrangian in PZW picture. However, the equivalence of the PZW picture and the gauge-invariant description with the minimal-coupling Lagrangian \labelcref{eq:ultimateAction} is manifested already here on the Lagrangian level. Since the description on the Lagrangian level does not involve choices of canonical momentum, furthermore, the two Lagrangian functionals \(L\) and \(L\PZW\) together with their variables, the field coordinates are gauge invariant, there cannot be such inconsistencies in the PZW picture as alleged in Ref.~\cite{Rousseau2017}.

\subsection{The Hamiltonian in the PZW picture}
\label{sec:pzwHamiltonian}

To construct the Hamiltonian in the PZW picture, we first have to determine the canonical momenta. The Lagrangian as a function of the generalized coordinates and velocities reads:
\begin{equation}
L\PZW(\vb x_\alpha,\dot{\vb x}_\alpha,\vb A\trans,\partial_t\vb A\trans) = \sum_{\alpha=1}^{Z} \, \frac{m_\alpha}{2} \, \dot{\rv}_\alpha^2 (t)
+ \spaceint \qty[ \frac{\epsNot}{2} \, \vb{E}^2  - \frac{1}{2 \, \mu_0} \, \vb{B}^2  ]
+ \spaceint \qty\big[\vb{P}  \cdot \vb{E}  + \vb{M}  \cdot \vb{B}  ] \; ,
\end{equation}
where we are again assuming the nucleus as immobile, this time also in the kinetic part. Here, on the left-hand-side, we again made explicit the fact that the field generalized coordinate remains \(\vb{A}\trans\) also in this picture, while the right-hand-side is written in a nice compact form with the physical fields, \(\vb B\) containing the field coordinate while \(\vb E\) the velocity (cf. \cref{eq:fieldsFromPotentials}).

\begin{figure}
 \includegraphics[width=\linewidth]{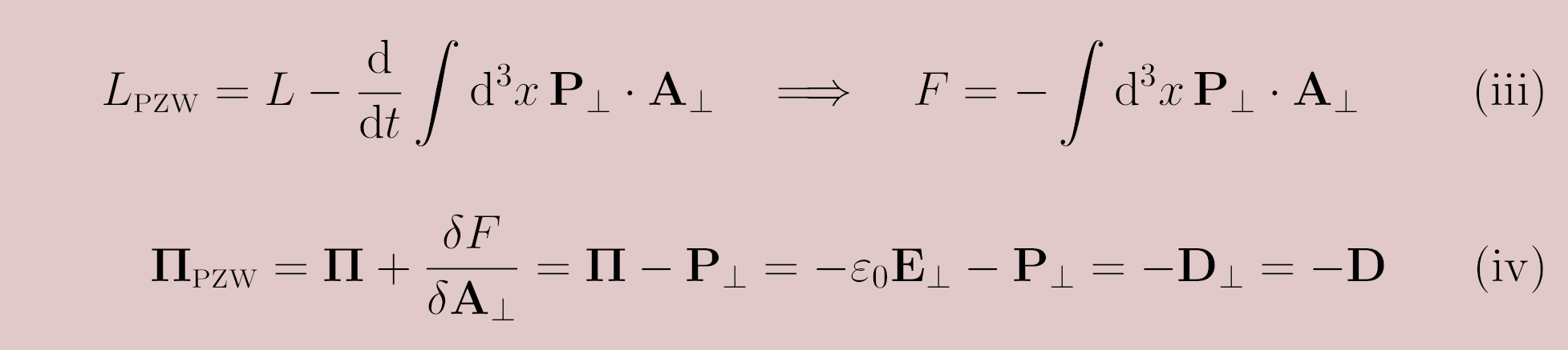}
 \caption{The change of conjugate momentum under the PZW transformation as derived from the general formula presented in \cref{fig:generalMomentumTrafo}}
 \label{fig:fieldMomentumTrafo}
\end{figure}

Canonical momenta are produced as functional derivatives of the Lagrangian along the generalized velocities:
\begin{subequations}
\begin{align}
\vb{p}_{\text{\tiny PZW},\alpha}  &= \fdv{L\PZW}{\dot{\rv}_\alpha}\; ,\\
\vb{\Pi}\PZW  &= \fdv{L\PZW}{\qty( \partial_t \, \vb{A}\trans  )} \; .
\end{align}
\end{subequations}
When evaluating the functional derivative along the particle velocity, we have to take care to differentiate also the term containing the magnetization. This term can be written in the following form:
\begin{equation}
\spaceint \vb{M}\cdot\vb{B}=-\sum_{\alpha=1}^{Z}q_\alpha\,\dot\rv_\alpha(t)\cdot
\qty[\rv_\alpha(t)\times
\sint \, s \, \vb{B} \qty( s \, \rv_\alpha (t), t  )].
\end{equation}
In this form, the variation along the particle velocity is easily performed to yield the magnetic contribution for the particle momentum, which reads
\begin{equation}
\label{eq:PZW_particleMomentum}
\vb{p}_{\text{\tiny PZW},\alpha} (t) = m_\alpha \, \dot{\rv}_\alpha (t) 
- q_\alpha \, \rv_\alpha (t)  \times
\sint \, s \, \vb{B} \qty( s \, \rv_\alpha (t), t  )
 \; .
\end{equation}
To calculate the field canonical momentum, let us identify the terms in \(L\PZW\) which contain \(\vb E\trans=-\partial_t\,\vb A\trans\):
\begin{equation}
\label{eq:PoinLagrangian}
 L\PZW=
 \qty(\parbox{6ex}{\scriptsize particle kinetic terms})+
 \qty(\parbox{7.5ex}{\scriptsize magnetic terms})+
 \spaceint\qty[\frac\epsNot2\qty(\vb E\longi^2+\vb E\trans^2)+\qty(\vb P\longi\cdot\vb E\longi+\vb P\trans\cdot\vb E\trans)],
\end{equation}
whence
\begin{equation}
\label{eq:PoinMomentumDerivation}
\vb{\Pi}\PZW=\fdv{L\PZW}{\qty( \partial_t \, \vb{A}\trans  )}=
 -\fdv{L\PZW}{\vb{E}\trans}=-\epsNot\vb E\trans-\vb P\trans=-\vb D\trans=-\vb D.
\end{equation}
The same result can be obtained by using the general formula of the change of momentum under an equivalent Lagrangian transformation, as we demonstrate in \cref{fig:fieldMomentumTrafo}.

Having the expressions for the momenta, we can finally determine the Hamiltonian in the PZW picture:
\begin{multline}
\label{eq:pzwHamiltonian}
H\PZW = \sum_{\alpha=1}^{Z} \vb{p}_{\text{\tiny PZW},\alpha}(t) \cdot \dot{\rv}_\alpha (t)
+ \spaceint \, \vb{\Pi}\PZW  \cdot \partial_t \, \vb{A}\trans  \, - L\PZW 
=\sum_{\alpha=1}^{Z} \, \frac{m_\alpha}{2} \, \dot{\rv}_\alpha^2 (t) 
+ \spaceint \qty( \frac{\epsNot}{2}\, \vb{E}^2  + \frac{1}{2 \, \mu_0} \, \vb{B}^2   ) \\
= \sum_{\alpha=1}^{Z} \, \frac{1}{2 \, m_\alpha} \,
\qty( \vb{p}_{\text{\tiny PZW},\alpha} (t) + q_\alpha \, \rv_\alpha (t)  \times
\sint \, s \, \vb{B} \qty( s \, \rv_\alpha (t) ) )^2  
+ \spaceint \qty( \frac{1}{2 \, \epsNot} \, \qty\big[ \vb{\Pi}\PZW  + \vb{P}  ]^2  
+ \frac{1}{2 \, \mu_0} \, \vb{B}^2   )\\=
\sum_{\alpha=1}^{Z} \, \frac{1}{2 \, m_\alpha} \,
\qty( \vb{p}_{\text{\tiny PZW},\alpha} (t) + + q_\alpha \, \rv_\alpha (t)  \times
\sint \, s \, \vb{B} \qty( s \, \rv_\alpha (t) ) )^2\\
+ \spaceint \qty( \frac{1}{2 \, \epsNot} \vb D^2 
+ \frac{1}{2 \, \mu_0} \, \vb{B}^2   )
- \frac 1\epsNot\spaceint\vb D\cdot\vb P
+ \frac1{2\,\epsNot}\spaceint \vb P^2\; ,
\end{multline}
where the field canonical coordinate is contained by \(\vb{B} = \curl{\vb{A}\PZW}\). After the last equality sign, we can discover the familiar PZW Hamiltonian, which
\begin{enumerate}
 \item is free from an electric A-square term (the magnetic contribution to the particle kinetic term is the so-called Röntgen term that vanishes in electric-dipole order),
 \item accounts for the light-matter interaction in the form of the \(\vb D\cdot\vb P\) term, and
 \item contains a P-square term, for which a straightforward regularization procedure was presented in \cite{Vukics15}.
\end{enumerate}

\section{Ambiguity of the field momenta}
\label{sec:momentumAmbiguity}

In this section, we show that even if the field generalized coordinate is fixed (say, to \(\vb A\trans\)), there is still a certain freedom in choosing the momentum conjugate to it, which freedom is of a field-theoretic nature. This is most easily shown in the case of the gauge-invariant Lagrangian \labelcref{eq:ultimateAction}. Using \cref{eq:dtA} we can identify that part of the Largangian which contains the field velocity:
\begin{equation}
 L=\spaceint \frac\epsNot2\,\vb{E}\trans^2+\text{…},
\end{equation}
and immediately derive the field momentum conjugate to the gauge-invariant coordinate \(\vb A\trans\) to obtain
\begin{equation}
\label{eq:naiveFieldMomentum}
\vb\Pi=\fdv{\Lcal_\text{field}}{\qty(\partial_t \vb A\trans)}=-\epsNot\,\vb E\trans.
\end{equation}

However, this part of the Lagrangian could be supplemented as
\begin{equation}
\label{eq:LagrangianSupplementTrick}
 L=\spaceint \qty(\frac\epsNot2\,\vb{E}\trans^2+\epsNot\,\vb{E}\longi\cdot\vb E\trans) + \text{…},
\end{equation}
since the spatial integral of the second term is zero. Varying this form, we find
\begin{multline}
\delta L=\spaceint\qty( \epsNot\,\vb{E}\trans\cdot\delta\vb{E}\trans+\epsNot\,\vb{E}\longi\cdot\delta\vb{E}\trans)=-\spaceint \epsNot\qty(\vb{E}\trans+\vb E\longi)\cdot\delta\qty(\partial_t\vb{A}\trans)\\=-\spaceint \epsNot\,\vb{E}\cdot\delta\qty(\partial_t\vb{A}\trans),
\end{multline}
yielding \(\vb \Pi=-\epsNot\,\vb E\) for the canonical field momentum. This is also the \emph{a priori} result from the generic Lagrangian \cref{eq:fundamentalAction}, when the variation is performed symbolically without regard to the interdependence of the variables. This is also one of the starting points of the paper by Rousseau and Felbacq \cite{Rousseau2017}. However, here we clarified that this is only one of several possible choices.

As explained by Weinberg in Section 11.3 of his book \cite{Weinberg}, the choice of \(\vb\Pi'=-\epsNot\,\vb E\) as canonical field momentum has the “awkward feature” (quote: Weinberg) that when quantized, it does not commute with the particle momenta, for the simple reason that \(\vb E\longi\) is determined by the particle coordinates. On the other hand, with the choice \(\vb\Pi=-\epsNot \,\vb{E}\trans\), the only nontrivial commutation relation will be the well-known (cf. Eq. (11.3.20) in \cite{Weinberg})
\begin{equation}
 \comm{\vb A\trans(\rv,t)}{\vb\Pi(\vb y,t)}=i\hbar\,\vb*\delta\trans\qty(\rv-\vb y).
\end{equation}
Here, \(\vb*\delta\trans\) is the transverse delta function, a 2nd-order tensor field.

Let us note that what we are doing here with \cref{eq:LagrangianSupplementTrick} is different from equivalent transformations of the Lagrangian (cf. \cref{eq:equivalentActionTrafo}), since here the change of the Lagrangian under the switch between the two possible choices of field momentum is effectively zero, so that for example the picture cannot change either.

In the following, we show that a similar freedom is present in the PZW picture.

\subsection{The connection with Poincaré gauge}
\label{sec:connectionPoincare}

The defining condition of Poincaré gauge reads:
\begin{equation}
\label{eq:PoincareGaugeDefiningCond}
\rv\cdot \vb{A}\Poin =0 \; .
\end{equation}
This condition is equivalent to a pair of expressions whereby the potentials are uniquely determined by the physical fields in this gauge:
\begin{subequations}
\label{eq:PoincareGaugePotentials}
\begin{align}
\label{eq:Phi_P}
\Phi\Poin  
&= \Phi_0 (t) - \rv \cdot \sint \, \vb{E} \qty( s \, \rv\,,t )  \; ,\\
\label{eq:A_P}
\vb{A}\Poin  &= - \rv \times 
\sint \, s \, \vb{B} \qty(s \, \rv\,,t )  \;  ,
\end{align}
\end{subequations}
where \(\Phi_0 (t)\) is an integration constant which, being independent of space, does not enter the physical fields.\footnote{Note that from \cref{eq:A_P}, the transverse vector potential determines the whole of \(\vb A\Poin\) by virtue of \(\vb B=\curl{\vb A}=\curl{\vb A\trans}\), which underlines that \(\vb A\trans\) is the true dynamical variable also in Poincaré gauge. We also note that the physical fields always uniquely determine the potentials with different expressions in any gauge.} It is clear that \cref{eq:PoincareGaugeDefiningCond} follows from \cref{eq:A_P}, while the other direction of the equivalence is proven in \Cref{app:PoincarePotentialsFromDefiningCond}.

Let us return to the “generic” formula of the Lagrangian \labelcref{eq:fundamentalAction}, and substitute the expressions \labelcref{eq:PoincareGaugePotentials}. The interaction terms then read:
\begin{subequations}
\begin{align}
\spaceint \, \rho  \, \Phi\Poin 
=& - \sum_{\alpha=1}^{Z} \, q_{\alpha} 
\,\rv_\alpha (t) \cdot \sint \, \vb{E}\qty( s \, \rv_\alpha (t)\,,t) & 
= && - \spaceint \, \vb{P}  \cdot \vb{E} \; ,\\
\spaceint \, \vb{j}  \cdot \vb{A}\Poin  
=& - \sum_{\alpha=1}^{Z} \, q_{\alpha} \; \dot{\rv}_\alpha (t)
\cdot \qty[ \rv_\alpha (t) \times \sint \, s \, \vb{B} \qty(s \, \rv_\alpha (t)\,,t ) ] & = &&\spaceint \, \vb{M}  \cdot \vb{B} \; ,
\end{align}
\end{subequations}
where we have used that the term containing \(\Phi_0 (t)\) vanishes due to the neutrality of the atom, while the second equality in each row comes from the comparison with \cref{eq:polarization,eq:magnetization}. This means that by expressing the interaction terms with the potentials in Poincaré gauge, we symbolically get the form \cref{eq:PZW_hatas} of the Lagrangian:
\begin{equation}
 L_\text{Poincaré}(\vb x_\alpha,\dot{\vb x}_\alpha,?)\overset{\text{symbolically}}{=}L\PZW(\vb x_\alpha,\dot{\vb x}_\alpha,\vb A\trans,\partial_t\vb A\trans)\,.
\end{equation}

What is important to note here is that though we have fixed the gauge in the Lagrangian \labelcref{eq:fundamentalAction} in the sense that we substituted the forms of the potentials in a certain gauge, we haven’t declared yet with what generalized coordinates we intend to describe the dynamics of the field. Gauge fixing and the choice of coordinates are two conceptually different things: the first fixes the expressions of the potentials, however, in the Lagrangian description of electrodynamics, nothing obliges us to use the (full) potentials as field coordinates!

The transverse part of the vector potential remains a good choice for the field coordinate also in Poincaré gauge. It is gauge invariant, i.e. \(\vb A\Poin\null_,\null\trans=\vb A\trans(=\vb A\Coul\null_,\null\trans(=\vb A\Coul))\), but, more importantly, it bijectively covers the c-manifold of the field. With this choice it is possible to say also in the physical sense that
\begin{equation}
 L_\text{Poincaré}(\vb x_\alpha,\dot{\vb x}_\alpha,\vb A\trans,\partial_t\vb A\trans)=L\PZW(\vb x_\alpha,\dot{\vb x}_\alpha,\vb A\trans,\partial_t\vb A\trans)\quad\text{for a single atom}.
\end{equation}
We see that even in this case, the equivalence of the Poincaré-gauge and the PZW Lagrangian comes with a qualification.\footnote{\label{fn:oneatom}Indeed, the PZW picture is easily extended to the case of several atoms = several charge centres, e.g. the polarization field can be chosen as \begin{equation}
\vb{P}  = \sum_A\sum_{\alpha\in A} q_{\alpha} \, \vb*{\xi}_{A,\alpha} (t) \, \sint \, \delta \qty( \rv - \rv\CoM\null_{,A} - s \, \vb*{\xi}_{A,\alpha} (t) ) \; ,
\end{equation}
where \(A\) indexes the different atoms – this form with separate charge centres for the different atoms being a convenient starting point for a multipolar approximation. But the Poincaré gauge defined by
\begin{equation}
 \qty(\rv-\rv\CoM)\cdot \vb{A}\Poin =0
\end{equation}
knows about a single centre of charge only. This only expresses the fact well-known in literature (cf. eg. \cite{CDG} Chapter IV.) that the set of equivalent transformations of the Lagrangian is wider than that of gauge transformations.
}

Let us now turn to the conjugate momentum in Poincaré gauge, where we can show that a similar ambiguity is present as noted above in the case of the gauge-invariant Lagrangian. As we have shown in \cref{eq:PoinMomentumDerivation}, the natural choice in Poincaré gauge is \(\vb\Pi\Poin=-\vb D\). However, using the identity
\begin{equation}
 \label{eq:magicIdentity}
 \spaceint \vb P\trans\cdot\vb E\trans=\spaceint \vb P\longi\cdot\qty(\partial_t\vb A\Poin)\longi,
\end{equation}
whose proof is given in \Cref{app:proof} to show that in the Lagrangian \labelcref{eq:PoinLagrangian}, one of the appearances of the field generalized velocity can be eliminated to give
\begin{equation}
\label{eq:PoinLagrangianModified}
 L\Poin=
 \qty(\parbox{6ex}{\scriptsize particle kinetic terms})+
 \qty(\parbox{7.5ex}{\scriptsize magnetic terms})+
 \spaceint\qty[\frac\epsNot2\qty(\vb E\longi^2+\vb E\trans^2)+\vb P\longi\cdot\qty(\vb E\longi+\qty(\partial_t\vb A\Poin)\longi)].
\end{equation}
Now the subscript \(\text{P}\) refers to either Poincaré gauge or PZW picture in the just discussed sense of equivalence between the two. What we have to note here is that \(\vb{A}\Poin\null_,\null\longi\) is not a field variable but one that belongs to the electrostatic part of the problem. This can be immediately seen from \cref{eq:fieldsFromPotentialsOne} because \(\vb E\trans\) is determined solely by \(\vb A\trans\) also in Poincaré gauge (both of these vector fields are in fact gauge invariant), while \(\Phi\Poin\) and \(\vb A\Poin\null_,\null\longi\) conspire to yield \(\vb E\longi\), which latter is determined by the particles (here \(\vb E\longi\) is gauge invariant, but \(\Phi\) and \(\vb A\longi\) are of course not). The derivation \labelcref{eq:PoinMomentumDerivation} is modified when we start out from the form \labelcref{eq:PoinLagrangianModified} to yield \(\vb \Pi\Poin'=-\epsNot\,\vb E\trans\) for the canonical field momentum.

How do we decide which momentum variable to use in this case? Here we cannot rely on the argument of convenience as in the case of the minimal-coupling picture above, since here the commutation relations will be the same with both choices: 
\begin{multline}
\label{eq:commutationPZW}
  \comm{\vb A\trans(\rv,t)}{-\vb D(\vb y,t)}=\comm{\vb A\trans(\rv,t)}{\vb\Pi\Poin(\vb y,t)}\\=i\hbar\,\vb*\delta\trans\qty(\rv-\vb y)\\=\comm{\vb A\trans(\rv,t)}{\vb\Pi\Poin'(\vb y,t)}=\comm{\vb A\trans(\rv,t)}{-\epsNot\vb E\trans(\vb y,t)}.
\end{multline}
The answer is that if we are aiming at the usual form of the PZW Hamiltonian \labelcref{eq:pzwHamiltonian}, then we have to choose \(-\vb D\) as the field canonical momentum because even though the equations of motion are independent, the \emph{form} of the Hamiltonian does depend on the choice of momentum.

When the full potentials \(\Phi\Poin\) and \(\vb A\Poin\) are chosen as field coordinate, the c-manifold is not bijectively covered since \(\Phi\Poin\) and part of \(\vb A\Poin\) belong to the electrostatic part of the problem, which is at the same time determined also by the particle coordinates.\footnote{In this case the coordinate space spanned by the \(\rv_\alpha\)s, \(\Phi\Poin\), and \(\vb A\Poin\) is “bigger” than the c-manifold because part of this coordinate space – where the particle coordinates, the scalar potential, and the longitudinal part of the vector potential determine the electrostatic part differently – is actually non-physical. This situation can be handled by the explicit use of the following constraint derived from \cref{eq:electrostaticGaugeInvariant}:\begin{equation}\partial_t\vb A\Poin\null_,\null\longi+\grad{\Phi\Poin}=\frac1{4\pi\epsNot}\sum_{\alpha=0}^Zq_\alpha\frac{\qty(\vb x_\alpha(t)-\vb x)}{\abs{\vb x_\alpha(t)-\vb x}^3}\;.\end{equation}} In this case the equivalence of the PZW picture with Poincaré gauge does not hold even in the relative sense described above:
\begin{equation}
 L_\text{Poincaré}(\vb x_\alpha,\dot{\vb x}_\alpha,\Phi\Poin,\vb A\Poin,\partial_t\vb A\Poin)\neq L\PZW(\vb x_\alpha,\dot{\vb x}_\alpha,\vb A\trans,\partial_t\vb A\trans),
\end{equation}
which can also be immediately seen because the domains of these two functionals are different.

\section{Summary}

In summary, we have shown that the Power–Zienau–Woolley picture can be derived from a gauge-invariant Lagrangian, in a way which does not make reference to any gauge or choice of canonical momentum. For a treatment emphasizing the unitary equivalence between the minimal-coupling and the PZW picture, cf. Ref.~\cite{Andrews2018}.

We believe that our analysis has clearly dissolved all the objections raised recently against the PZW picture by Rousseau and Felbacq \cite{Rousseau2017}. In the following, we briefly react to some central claims of theirs which appear erroneous in the light of our treatment.
\begin{enumerate}
 \item Talking about a “multipolar gauge” is strictly speaking not correct, and this is not how the PZW picture was understood in the literature, either \cite{Andrews2018}. As we have seen in \Cref{sec:connectionPoincare}, it is only in a strongly qualified sense that we can talk about the equivalence of the Poincaré gauge and the PZW picture (e.g. only in the case of a single charge centre, that is, a single atom), but such an idea can only occur if the transverse vector potential is used as field coordinate irrespective of gauge, as is the case in the PZW picture.
 \item The PZW picture cannot be declared inconsistent on the basis that it is not derived via a gauge transformation. Here we have shown that the minimal-coupling picture can be formulated in a gauge-invariant manner, and the PZW picture is equivalent to this in the sense that it can be derived through an equivalent Lagrangian transformation regardless of gauge. It is furthermore a well-known fact that such transformations are more general than gauge transformations. However, the field coordinate – which is the gauge invariant \(\vb A\trans\) in both pictures – remains untouched by a Lagrangian transformation, as this is part of the definition of the Lagrangian problem in the first place. Moreover, both the Lagrangian \labelcref{eq:PZW_hatas} and the Hamiltonian \labelcref{eq:pzwHamiltonian} can be expressed solely with the physical fields in the PZW picture. When the PZW Hamiltonian is used in Schrödinger picture, as is often the case in quantum optics e.g. in the derivation of the Jaynes-Cummings model, then potentials do not play a role at all, further emphasizing the irrelevance of gauge.
 \item It is incorrect to expect that the canonical momentum is gauge invariant, since the momentum in general does change under an equivalent Lagrangian transformation in the form of \cref{eq:equivalentActionTrafo} as demonstrated in \cref{fig:generalMomentumTrafo,fig:fieldMomentumTrafo}. Sure, the electric field \(\vb E\) is gauge invariant, \emph{but its capacity of being the canonical field momentum is not}.
 \item The appearance of the displacement field \(\vb D\) in the PZW picture does not mean that concepts from macroscopic electrodynamics are mixed into the electrodynamics of atoms. The replacement of the charge and current densities with polarization and magnetization densities is just another way of describing the same things.
 \item It is incorrectly argued that the A-square term is present in the PZW picture. It is true that in the particle momentum \labelcref{eq:PZW_particleMomentum}, the magnetic contribution is exactly \(\vb A\Poin\), but this is still a magnetic term, which will be neglected in electric-dipole order, so that it does not cause the same problems as the “electric” A-square term in Coulomb gauge.
\end{enumerate}
In Ref.~\cite{Rousseau2017}, remarkable effort is taken to calculate the Dirac brackets in the case when the full \(\vb A\Poin\) is taken as field coordinate and \(-\epsNot\vb E\) as conjugate momentum. This is, however, an unnecessary complication, in analogy with the choice of \(\vb A\trans\) and the full \(-\epsNot\vb E\) in the minimal-coupling picture, as discussed in \Cref{sec:momentumAmbiguity}. In our treatment, in the PZW picture just like in the minimal-coupling one the only non-trivial Dirac bracket is the one between the field coordinate and conjugate momentum, and is proportional to \(\vb*\delta\trans\) (cf. \cref{eq:commutationPZW}).

{\small
\subsubsection*{Contributions}

P.D. initiated this work. G.K. and A.V. did the calculations. All three authors discussed the results and interpretations together.

\subsubsection*{Competing Interests}

The authors declare that they have no competing interests.

\subsubsection*{Data Availability}

No datasets were generated or analysed during the current study.}

\begin{acknowledgments}
We acknowledge valuable exchange with R. G. Woolley, E. Rousseau, and D. Felbacq. This work was supported by the National Research, Development and Innovation Office of Hungary (NKFIH) within the Quantum Technology National Excellence Program  (Project No. 2017-1.2.1-NKP-2017-00001) and by Grant No. K115624. A. V. acknowledges support from the János Bolyai Research Scholarship of the Hungarian Academy of Sciences.
\end{acknowledgments}

\appendix

\section{Polarization and magnetization fields from distribution theory}
\label{app:distros}
Using the neutrality of the atom, it is convenient to derive the charge and current density fields from polarization and magnetization fields.

Let us start with the charge density. Since in the case of point charges this field is a distribution, we will treat it in a distribution theoretic approach, that is, in integral form together with an appropriate test function \(f(\rv)\), which is smooth and vanishes in the spatial infinity:
\begin{equation}
\label{eq:chargeDistroApp}
\spaceint \, \rho(\rv,t) \, f(\rv) = \sum_{\alpha=0}^{Z} q_{\alpha} \, f \qty( \rv\CoM + \vb*{\xi}_\alpha (t) ) \; .
\end{equation}
Let us draw a line between the points \(\rv\CoM\) and \(\rv\CoM + \vb*{\xi}_\alpha (t)\). The points of the line segment between these two points can be written as \(\rv\CoM + s \, \vb*{\xi}_\alpha (t)\), where \(s \in \qty[0,1]\). Then, the following transformation can be performed:
\begin{equation}
f \qty( \rv\CoM + \vb*{\xi}_\alpha (t) ) = f \qty( \rv\CoM ) 
+ \sint \, \frac{d}{ds} f \qty( \rv\CoM + s \, \vb*{\xi}_\alpha (t) ) 
= f \qty( \rv\CoM ) 
+ \sint \; \vb*{\xi}_\alpha (t) \cdot \grad{f \qty( \rv\CoM + s \, \vb*{\xi}_\alpha (t) ) } \; .
\end{equation}
Let us write this back into the formula \cref{eq:chargeDistroApp} of the charge density. Due to the neutrality of the atom, the first term vanishes to yield
\begin{equation}
\spaceint \, \rho(\rv,t) \, f(\rv) = \sum_{\alpha=0}^{Z} q_{\alpha} \, \sint \, \vb*{\xi}_\alpha (t) \cdot \grad{ f \qty( \rv\CoM + s \, \vb*{\xi}_\alpha (t) ) }\; .
\end{equation}
The value of the gradient of the function at the points of the line segment between the two points can be represented as:
\begin{equation}
\label{eq:gradWithTranslatedDelta}
\grad{ f \qty( \rv\CoM + s \, \vb*{\xi}_\alpha (t) ) }
= \spaceint \, \grad{ f (\rv) }
\cdot \delta \qty( \rv - \rv\CoM - s \, \vb*{\xi}_\alpha (t) ) 
= - \spaceint \, f(\rv) 
\cdot \grad{ \delta \qty( \rv - \rv\CoM - s \, \vb*{\xi}_\alpha (t) ) }\; .
\end{equation}
Writing this back:
\begin{equation}
\spaceint \, \rho(\rv,t) \, f(\rv) = - \spaceint \, f(\rv) \, \sum_{\alpha=0}^{Z} q_{\alpha} \, \vb*{\xi}_\alpha (t)
\cdot \sint \, \bs{\nabla} \, \delta \qty( \rv - \rv\CoM - s \, \vb*{\xi}_\alpha (t) ) \; .
\end{equation}
Let us introduce the \emph{polarization field} with the following definition:
\begin{equation}
\vb{P} (\rv,t) = \sum_{\alpha=0}^{Z} q_{\alpha} \, \vb*{\xi}_\alpha (t) \, \sint \, \delta \qty( \rv - \rv\CoM - s \, \vb*{\xi}_\alpha (t)  ) \; .
\end{equation}
Then
\begin{equation}
\spaceint \, \rho(\rv,t) \, f(\rv) = - \spaceint \, f(\rv) \, \div\vb{P} (\rv,t) \; ,
\end{equation}
and since this is true for an arbitrary test function, we can deduce the following identity in the distribution theoretic sense:
\begin{equation}
\rho = - \div{\vb{P}} \; ,
\end{equation}
meaning that we have managed to derive the charge density from a polarization field.

In the next step, we want to do something analogous with the current density field as well. Let us start with the continuity equation and substitute therein the above formula of for the charge density. We get:
\begin{equation}
\div( \vb{j} - \partial_t \vb{P} ) = 0 \; ,
\end{equation}
that is, the field in the parentheses is transverse. Let us integrate it together with a distribution theoretic test function:
\begin{multline}
\spaceint \, \qty[ \vb{j} (\rv,t) - \partial_t \vb{P}(\rv,t) ]\, f(\rv) \\= \sum_{\alpha=0}^{Z} q_{\alpha} \, \dot{\vb*{\xi}}_{\alpha} (t) \, f \qty( \rv\CoM + \vb*{\xi}_\alpha (t) ) - \partial_t\qty[\sum_{\alpha=0}^{Z} q_{\alpha} \, \vb*{\xi}_{\alpha} (t) \, \sint \, f \qty( \rv\CoM + s \, \vb*{\xi}_\alpha (t) ) ] \; .
\end{multline}
Let us perform the time differentiation:
\begin{multline}
\spaceint \, \qty[ \vb{j} (\rv,t) - \partial_t \vb{P}\,  (\rv,t) ]\, f(\rv) = \sum_{\alpha=0}^{Z} q_{\alpha} \, \dot{\vb*{\xi}}_{\alpha} (t) \, f \qty( \rv\CoM + \vb*{\xi}_\alpha (t) ) \\
- \sum_{\alpha=0}^{Z} q_{\alpha} \, \dot{\vb*{\xi}}_{\alpha} (t) \, \sint \, f \qty( \rv\CoM + s \, \vb*{\xi}_\alpha (t) )
- \sum_{\alpha=0}^{Z} q_{\alpha} \, \vb*{\xi}_{\alpha} (t) \, \sint \, s \, \qty(\dot{\vb*{\xi}}_{\alpha} (t) \cdot \grad) f \qty( \rv\CoM + s \, \vb*{\xi}_\alpha (t) ) \; .
\end{multline}
In the second term on the right-hand side, we perform an integration by parts along \(s\):
\begin{multline}
\sint \, 1 \cdot f \qty( \rv\CoM + s \, \vb*{\xi}_\alpha (t) )
= \eval{s \, f \qty( \rv\CoM + s \, \vb*{\xi}_\alpha (t) ) }_{s=0}^{1}
- \sint \, s \cdot \dv{f}{s}\qty( \rv\CoM + s \, \vb*{\xi}_\alpha (t) ) \\
= f \qty( \rv\CoM + \vb*{\xi}_\alpha (t) )
- \sint  s \, \qty(\vb*{\xi}_{\alpha} (t) \cdot \grad )\, f \qty( \rv\CoM + s \, \vb*{\xi}_\alpha (t) ) \; .
\end{multline}
Writing this result back into the above formula, we notice that the first term is eliminated, and the remaining two can be contracted to give:
\begin{multline}
\spaceint \, \qty[ \vb{j} (\rv,t) - \partial_t \vb{P}(\rv,t) ]\, f(\rv) 
= \sum_{\alpha=0}^{Z} q_{\alpha} \qty[ \dot{\vb*{\xi}}_{\alpha} (t) \otimes \vb*{\xi}_{\alpha} (t) - \vb*{\xi}_{\alpha} (t) \otimes \dot{\vb*{\xi}}_{\alpha} (t) ] \sint \, s \, \grad f \qty( \rv\CoM + s \, \vb*{\xi}_\alpha (t) ) \\
= \spaceint \, f(\rv) \, \sum_{\alpha=0}^{Z} q_{\alpha} \qty[ \vb*{\xi}_{\alpha} (t) \otimes \dot{\vb*{\xi}}_{\alpha} (t) - \dot{\vb*{\xi}}_{\alpha} (t) \otimes \vb*{\xi}_{\alpha} (t) ] \sint \, s \, \grad\delta \qty( \rv - \rv\CoM - s \, \vb*{\xi}_\alpha (t) )\\
= \spaceint \, f(\rv) \, \sum_{\alpha=0}^{Z} q_{\alpha} \,
\qty[\sint \, s \, \grad \delta \qty( \rv - \rv\CoM - s \, \vb*{\xi}_\alpha (t) ) ]\times\qty( \vb*{\xi}_\alpha  \times \dot{\vb*{\xi}}_\alpha )\; .
\end{multline}
Where for the second equality, we have used \cref{eq:gradWithTranslatedDelta}, and for the third, the well-known identity of the vector triple product. Let us introduce the magnetization field as follows:
\begin{equation}
\vb{M} (\rv,t) = \sum_{\alpha=0}^{Z} q_{\alpha} \, \vb*{\xi}_\alpha (t) \times \dot{\vb*{\xi}}_\alpha (t) \,
\sint \, s \, \delta \qty( \rv - \rv\CoM - s \, \vb*{\xi}_\alpha (t)  ) \; .
\end{equation}
Hence
\begin{equation}
\spaceint \, \qty[ \vb{j}(\rv,t) - \partial_t \vb{P}\,  (\rv,t) ]\, f(\rv) = \spaceint \, f(\rv) \, \curl{\vb{M}(\rv,t)} \; ,
\end{equation}
from which, being true for an arbitrary test function, we can deduce the equality in the distribution theoretic sense:
\begin{equation}
\vb{j} = \frac{\partial \vb{P}}{\partial t} + \curl{\vb{M}} \; .
\end{equation}
We can see hence that the current density is composed of two parts, one being related with the electric polarizability and the other with the magnetizability of the atom.

\section{Potentials from the condition defining the Poincaré gauge}
\label{app:PoincarePotentialsFromDefiningCond}

For the definition of the Poincaré gauge, we must first choose a point of reference, which we denote by \(\rv_0\). At the point of reference, the value of the scalar potential at any time instant can be freely prescribed, let this be \(\Phi_0 (t)\). The condition defining the Poincaré gauge reads
\begin{equation}
\label{eq:Poincare_mertek}
\qty( \rv - \rv_0 ) \cdot \vb{A}\Poin (\rv,t) =0 \; .
\end{equation}
What we are proving in this Appendix is that the gauge fixing condition and the prescribed value of the scalar potential completely determine the potentials.

These two conditions can be automatically satisfied by looking for the potentials in the following form:
\begin{subequations}
\label{eq:potencialok_uvbol}
\begin{align}
\label{eq:Phi_P_u_bol}
\Phi\Poin (\rv,t) &= \Phi_0 (t) - \qty( \rv - \rv_0 ) \cdot \vb{u}(\rv,t) \; ,\\
\label{eq:A_P_v_bol}
\vb{A}\Poin (\rv,t)& = - \qty( \rv - \rv_0 )  \times \vb{v} (\rv,t) \; ,
\end{align}
\end{subequations}
where \(\vb{u} (\rv,t)\) and \(\vb{v} (\rv,t)\) are auxiliary vector fields. The potentials do not uniquely determine the auxiliary vector fields, as the latter can be transformed in the following way without changing the potentials:
\begin{subequations}
\begin{align}
\vb{u}' (\rv,t) &= \vb{u} (\rv,t) + \qty( \rv - \rv_0 ) \times \vb{w} (\rv,t) \; ,\\
\vb{v}' (\rv,t) &= \vb{v} (\rv,t) + \qty( \rv - \rv_0 ) \cdot \varphi (\rv,t) \; .
\end{align}
\end{subequations} 
In the following, this freedom in the auxiliary vector fields will be eliminated by prescribing extra conditions. Since we strive to relate \(\vb{u}\) with \(\vb{E}\) and \(\vb{v}\) with \(\vb{B}\), these extra conditions are fashioned after Maxwell’s homogeneous equations as
\begin{subequations}
\label{eq:uv_onkenyes_feltetel}
\begin{align}
\label{eq:u_onkenyes_feltetel}
\curl \vb{u}& = - \partial_t \, \vb{v} \; ,\\
\label{eq:v_onkenyes_feltetel}
\div \vb{v} &= 0 \; .
\end{align}
\end{subequations}  

In the next step, let us see what equations we get for \(\vb{u}\) and \(\vb{v}\) if the formulas \cref{eq:potencialok_uvbol} are substituted into \cref{eq:fieldsFromPotentials}. To simplify notation, we choose the origin as point of reference: \(\rv_0=\vb{0}\).
Then:
\begin{subequations}
\begin{equation}
\vb{E} = \rv \times \partial_t \vb{v} 
+ \grad \qty( \rv \cdot \vb{u} ) \; ,
\end{equation}
\begin{equation}
\vb{B} = -\curl \qty( \rv \times \vb{v} ) \; .
\end{equation}
\end{subequations}
Using \cref{eq:u_onkenyes_feltetel} and expanding the vector triple products, we get after simplification (using also \cref{eq:v_onkenyes_feltetel} in the expression of \(\vb B\)):
\begin{subequations}
\begin{align}
\vb{E} (\rv,t) &= \bigl( \rv \cdot \grad \bigr) \, \vb{u} (\rv,t) + \vb{u} (\rv,t)  \; ,\\
\vb{B} (\rv,t) &= \bigl( \rv \cdot \grad \bigr) \, \vb{v} (\rv,t) + 2 \, \vb{v} (\rv,t)   \; .
\end{align}
\end{subequations}
So, these two equations are to be solved for the auxiliary vector fields.

Let us perform the substitution \(\rv\to s \, \rv\), where \(s\) is a scalar parameter:
\begin{subequations}
\begin{align}
\vb{E} (s \, \rv,t) &= s \, \bigl( \rv \cdot \grad \bigr) \, \vb{u} (s \, \rv,t) + \vb{u} (s \, \rv,t)  \; ,\\
\label{eq:Beq_special}
\vb{B} (s \, \rv,t) &= s \, \bigl( \rv \cdot \grad \bigr) \, \vb{v} (s \, \rv,t) + 2 \, \vb{v} (s \, \rv,t)   \; .
\end{align}
\end{subequations}
The introduction of this parameter is useful as we can notice that
\begin{equation}
\dv{s} \, \vb{u} (s \, \rv,t) = \bigl( \rv \cdot \grad \bigr) \, \vb{u} (s \, \rv,t)  \; ,\quad\text{the same being true for }\vb v,
\end{equation}
so that the equations become ordinary differential equations along \(s\). Let us multiply \cref{eq:Beq_special} with \(s\) and contract the terms on the right hand side:
\begin{subequations}
\begin{align}
\vb{E} (s \, \rv,t) &= \dv{s} \biggl( s \, \vb{u} (s \, \rv,t) \biggr) \; ,\\
s \, \vb{B} (s \, \rv,t) &= \dv{s} \biggl( s^2 \, \vb{v} (s \, \rv,t) \biggr)   \; .
\end{align}
\end{subequations}
Let us integrate these two equations along \(s\) from \(0\) to \(1\):
\begin{subequations}
\label{eq:uv_vegeredmeny}
\begin{align}
\label{eq:u_vegeredmeny}
\sint \vb{E} (s \, \rv,t) &= \eval{s \, \vb{u} (s \, \rv,t) }_{s=0}^{1}
= \vb{u} (\rv,t) \; ,\\
\label{eq:v_vegeredmeny}
\sint s \, \vb{B} (s \, \rv,t) &= \eval{ s^2 \, \vb{v} (s \, \rv,t) }_{s=0}^{1} 
= \vb{v} (\rv,t)  \; .
\end{align}
\end{subequations}
Whereby we have determined the two auxiliary vector fields and hence the potentials in Poincaré gauge. The only remaining task is to verify that the solutions \cref{eq:uv_vegeredmeny} satisfy the conditions \cref{eq:uv_onkenyes_feltetel}, as this is obviously required for the consistency of our solution. But the satisfaction of these extra conditions follows directly from the homogeneous Maxwell equations.

\section{Proof of the identity \labelcref{eq:magicIdentity}}
\label{app:proof}
Using the expressions \cref{eq:polarization} and \cref{eq:A_P}, we can derive
\begin{multline}
\label{eq:magicIdentityProof}
 \spaceint\vb P\cdot\qty(\partial_t{\vb A}\Poin)=\spaceint
 \qty[\sum_{\alpha=1}^{Z} q_{\alpha} \, \rv_\alpha (t) \, \sint \, \delta \qty( \rv - s \, \rv_\alpha (t)  )]\cdot
 \qty[- \rv \times \int_0^1\dd{s'} \, s' \,\qty(\partial_t\vb{B} \qty(s' \, \rv\,,t )) ]\\=-\sum_{\alpha=1}^{Z}q_\alpha\sint\int_0^1\dd{s'}\,s'\;\rv_\alpha(t)\cdot\qty[\spaceint\,\delta \qty( \rv - s \, \rv_\alpha (t) )\;\rv\times\qty(\partial_t\vb{B} \qty(s' \, \rv\,,t ))]\\=-\sum_{\alpha=1}^{Z}q_\alpha\sint s\int_0^1\dd{s'}\,s'\;\rv_\alpha(t)\cdot\qty[\rv_\alpha(t)\times\qty(\partial_t\vb{B} \qty(s\,s' \, \rv_\alpha(t)\,,t ))],
\end{multline}
where the scalar triple product vanishes, due to the equality of two of its vectors. Then, using the gauge invariance of the transverse part of the vector potential, we have
\begin{multline}
 0=\spaceint\vb P\cdot\qty(\partial_t{\vb A}\Poin)=\spaceint\vb P\longi\cdot\qty(\partial_t{\vb A}\Poin)\longi+\spaceint\vb P\trans\cdot\qty(\partial_t{\vb A}\Poin)\trans\\=\spaceint\vb P\longi\cdot\qty(\partial_t{\vb A}\Poin)\longi-\spaceint\vb P\trans\cdot\vb E\trans,
\end{multline}
which is just \cref{eq:magicIdentity}.


\begin{thebibliography}{16}%
\makeatletter
\providecommand \@ifxundefined [1]{%
 \@ifx{#1\undefined}
}%
\providecommand \@ifnum [1]{%
 \ifnum #1\expandafter \@firstoftwo
 \else \expandafter \@secondoftwo
 \fi
}%
\providecommand \@ifx [1]{%
 \ifx #1\expandafter \@firstoftwo
 \else \expandafter \@secondoftwo
 \fi
}%
\providecommand \natexlab [1]{#1}%
\providecommand \enquote  [1]{``#1''}%
\providecommand \bibnamefont  [1]{#1}%
\providecommand \bibfnamefont [1]{#1}%
\providecommand \citenamefont [1]{#1}%
\providecommand \href@noop [0]{\@secondoftwo}%
\providecommand \href [0]{\begingroup \@sanitize@url \@href}%
\providecommand \@href[1]{\@@startlink{#1}\@@href}%
\providecommand \@@href[1]{\endgroup#1\@@endlink}%
\providecommand \@sanitize@url [0]{\catcode `\\12\catcode `\$12\catcode
  `\&12\catcode `\#12\catcode `\^12\catcode `\_12\catcode `\%12\relax}%
\providecommand \@@startlink[1]{}%
\providecommand \@@endlink[0]{}%
\providecommand \url  [0]{\begingroup\@sanitize@url \@url }%
\providecommand \@url [1]{\endgroup\@href {#1}{\urlprefix }}%
\providecommand \urlprefix  [0]{URL }%
\providecommand \Eprint [0]{\href }%
\providecommand \doibase [0]{http://dx.doi.org/}%
\providecommand \selectlanguage [0]{\@gobble}%
\providecommand \bibinfo  [0]{\@secondoftwo}%
\providecommand \bibfield  [0]{\@secondoftwo}%
\providecommand \translation [1]{[#1]}%
\providecommand \BibitemOpen [0]{}%
\providecommand \bibitemStop [0]{}%
\providecommand \bibitemNoStop [0]{.\EOS\space}%
\providecommand \EOS [0]{\spacefactor3000\relax}%
\providecommand \BibitemShut  [1]{\csname bibitem#1\endcsname}%
\let\auto@bib@innerbib\@empty
\bibitem [{\citenamefont {Rousseau}\ and\ \citenamefont
  {Felbacq}(2017)}]{Rousseau2017}%
  \BibitemOpen
  \bibfield  {author} {\bibinfo {author} {\bibfnamefont {E.}~\bibnamefont
  {Rousseau}}\ and\ \bibinfo {author} {\bibfnamefont {D.}~\bibnamefont
  {Felbacq}},\ }\href {\doibase 10.1038/s41598-017-11076-5} {\bibfield
  {journal} {\bibinfo  {journal} {Scientific Reports}\ }\textbf {\bibinfo
  {volume} {7}},\ \bibinfo {pages} {11115} (\bibinfo {year}
  {2017})}\BibitemShut {NoStop}%
\bibitem [{\citenamefont {Power}\ and\ \citenamefont
  {Zienau}(1959)}]{Power1959427}%
  \BibitemOpen
  \bibfield  {author} {\bibinfo {author} {\bibfnamefont {E.}~\bibnamefont
  {Power}}\ and\ \bibinfo {author} {\bibfnamefont {S.}~\bibnamefont {Zienau}},\
  }\href@noop {} {\bibfield  {journal} {\bibinfo  {journal} {Philos. Trans. R.
  Soc. London, Ser. A}\ }\textbf {\bibinfo {volume} {251}},\ \bibinfo {pages}
  {427} (\bibinfo {year} {1959})}\BibitemShut {NoStop}%
\bibitem [{\citenamefont {Woolley}(1971)}]{Woolley1971}%
  \BibitemOpen
  \bibfield  {author} {\bibinfo {author} {\bibfnamefont {R.}~\bibnamefont
  {Woolley}},\ }\href@noop {} {\bibfield  {journal} {\bibinfo  {journal} {Proc.
  R. Soc. London, Ser. A}\ }\textbf {\bibinfo {volume} {321}},\ \bibinfo
  {pages} {1547} (\bibinfo {year} {1971})}\BibitemShut {NoStop}%
\bibitem [{\citenamefont {Woolley}(1974)}]{Woolley1974488}%
  \BibitemOpen
  \bibfield  {author} {\bibinfo {author} {\bibfnamefont {R.}~\bibnamefont
  {Woolley}},\ }\href {\doibase 10.1088/0022-3700/7/4/023} {\bibfield
  {journal} {\bibinfo  {journal} {Journal of Physics B: Atomic and Molecular
  Physics}\ }\textbf {\bibinfo {volume} {7}},\ \bibinfo {pages} {488} (\bibinfo
  {year} {1974})}\BibitemShut {NoStop}%
\bibitem [{\citenamefont {Woolley}(1980)}]{Woolley19802795}%
  \BibitemOpen
  \bibfield  {author} {\bibinfo {author} {\bibfnamefont {R.}~\bibnamefont
  {Woolley}},\ }\href {\doibase 10.1088/0305-4470/13/8/027} {\bibfield
  {journal} {\bibinfo  {journal} {Journal of Physics A: Mathematical and
  General}\ }\textbf {\bibinfo {volume} {13}},\ \bibinfo {pages} {2795}
  (\bibinfo {year} {1980})}\BibitemShut {NoStop}%
\bibitem [{\citenamefont {Babiker}\ and\ \citenamefont
  {Loudon}(1983)}]{Babiker1983}%
  \BibitemOpen
  \bibfield  {author} {\bibinfo {author} {\bibfnamefont {M.}~\bibnamefont
  {Babiker}}\ and\ \bibinfo {author} {\bibfnamefont {R.}~\bibnamefont
  {Loudon}},\ }\href@noop {} {\bibfield  {journal} {\bibinfo  {journal} {Proc.
  R. Soc. London, Ser. A}\ }\textbf {\bibinfo {volume} {385}},\ \bibinfo
  {pages} {1789} (\bibinfo {year} {1983})}\BibitemShut {NoStop}%
\bibitem [{\citenamefont {Power}\ and\ \citenamefont
  {Thirunamachandran}(1983)}]{Power19832649}%
  \BibitemOpen
  \bibfield  {author} {\bibinfo {author} {\bibfnamefont {E.}~\bibnamefont
  {Power}}\ and\ \bibinfo {author} {\bibfnamefont {T.}~\bibnamefont
  {Thirunamachandran}},\ }\href {\doibase 10.1103/PhysRevA.28.2649} {\bibfield
  {journal} {\bibinfo  {journal} {Physical Review A}\ }\textbf {\bibinfo
  {volume} {28}},\ \bibinfo {pages} {2649} (\bibinfo {year}
  {1983})}\BibitemShut {NoStop}%
\bibitem [{\citenamefont {Ackerhalt}\ and\ \citenamefont
  {Milonni}(1984)}]{Ackerhalt1984116}%
  \BibitemOpen
  \bibfield  {author} {\bibinfo {author} {\bibfnamefont {J.}~\bibnamefont
  {Ackerhalt}}\ and\ \bibinfo {author} {\bibfnamefont {P.}~\bibnamefont
  {Milonni}},\ }\href {\doibase 10.1364/JOSAB.1.000116} {\bibfield  {journal}
  {\bibinfo  {journal} {Journal of the Optical Society of America B: Optical
  Physics}\ }\textbf {\bibinfo {volume} {1}},\ \bibinfo {pages} {116} (\bibinfo
  {year} {1984})}\BibitemShut {NoStop}%
\bibitem [{\citenamefont {Vukics}\ \emph {et~al.}(2014)\citenamefont {Vukics},
  \citenamefont {Grießer},\ and\ \citenamefont {Domokos}}]{Vukics2014}%
  \BibitemOpen
  \bibfield  {author} {\bibinfo {author} {\bibfnamefont {A.}~\bibnamefont
  {Vukics}}, \bibinfo {author} {\bibfnamefont {T.}~\bibnamefont {Grießer}}, \
  and\ \bibinfo {author} {\bibfnamefont {P.}~\bibnamefont {Domokos}},\ }\href
  {\doibase 10.1103/PhysRevLett.112.073601} {\bibfield  {journal} {\bibinfo
  {journal} {Physical Review Letters}\ }\textbf {\bibinfo {volume} {112}},\
  \bibinfo {pages} {073601} (\bibinfo {year} {2014})}\BibitemShut {NoStop}%
\bibitem [{\citenamefont {Cohen-Tannoudji}\ \emph {et~al.}(1997)\citenamefont
  {Cohen-Tannoudji}, \citenamefont {Dupont-Roc},\ and\ \citenamefont
  {Grynberg}}]{CDG}%
  \BibitemOpen
  \bibfield  {author} {\bibinfo {author} {\bibfnamefont {C.}~\bibnamefont
  {Cohen-Tannoudji}}, \bibinfo {author} {\bibfnamefont {J.}~\bibnamefont
  {Dupont-Roc}}, \ and\ \bibinfo {author} {\bibfnamefont {G.}~\bibnamefont
  {Grynberg}},\ }\href@noop {} {\emph {\bibinfo {title} {Photons and Atoms}}}\
  (\bibinfo  {publisher} {Wiley-Interscience},\ \bibinfo {year}
  {1997})\BibitemShut {NoStop}%
\bibitem [{\citenamefont {Milonni}(1994)}]{Milonni1994}%
  \BibitemOpen
  \bibfield  {author} {\bibinfo {author} {\bibfnamefont {P.}~\bibnamefont
  {Milonni}},\ }\href@noop {} {\emph {\bibinfo {title} {The Quantum Vacuum}}}\
  (\bibinfo {year} {1994})\BibitemShut {NoStop}%
\bibitem [{\citenamefont {Compagno}\ \emph {et~al.}(1995)\citenamefont
  {Compagno}, \citenamefont {Passante},\ and\ \citenamefont
  {Persico}}]{Compagno1995}%
  \BibitemOpen
  \bibfield  {author} {\bibinfo {author} {\bibfnamefont {G.}~\bibnamefont
  {Compagno}}, \bibinfo {author} {\bibfnamefont {R.}~\bibnamefont {Passante}},
  \ and\ \bibinfo {author} {\bibfnamefont {F.}~\bibnamefont {Persico}},\
  }\href@noop {} {\emph {\bibinfo {title} {Atom-Field Interactions and Dressed
  Atoms}}}\ (\bibinfo {year} {1995})\BibitemShut {NoStop}%
\bibitem [{\citenamefont {Landau}\ and\ \citenamefont
  {Lifshitz}(1976)}]{landau}%
  \BibitemOpen
  \bibfield  {author} {\bibinfo {author} {\bibfnamefont {L.~D.}\ \bibnamefont
  {Landau}}\ and\ \bibinfo {author} {\bibfnamefont {E.~M.}\ \bibnamefont
  {Lifshitz}},\ }\href@noop {} {\emph {\bibinfo {title} {Mechanics}}}\
  (\bibinfo  {publisher} {Butterworth–Heinemann},\ \bibinfo {year}
  {1976})\BibitemShut {NoStop}%
\bibitem [{\citenamefont {Vukics}\ \emph {et~al.}(2015)\citenamefont {Vukics},
  \citenamefont {Grie\ss{}er},\ and\ \citenamefont {Domokos}}]{Vukics15}%
  \BibitemOpen
  \bibfield  {author} {\bibinfo {author} {\bibfnamefont {A.}~\bibnamefont
  {Vukics}}, \bibinfo {author} {\bibfnamefont {T.}~\bibnamefont {Grie\ss{}er}},
  \ and\ \bibinfo {author} {\bibfnamefont {P.}~\bibnamefont {Domokos}},\ }\href
  {\doibase 10.1103/PhysRevA.92.043835} {\bibfield  {journal} {\bibinfo
  {journal} {Phys. Rev. A}\ }\textbf {\bibinfo {volume} {92}},\ \bibinfo
  {pages} {043835} (\bibinfo {year} {2015})}\BibitemShut {NoStop}%
\bibitem [{\citenamefont {Weinberg}(2013)}]{Weinberg}%
  \BibitemOpen
  \bibfield  {author} {\bibinfo {author} {\bibfnamefont {S.}~\bibnamefont
  {Weinberg}},\ }\href@noop {} {\emph {\bibinfo {title} {Lectures on Quantum
  Mechanics}}}\ (\bibinfo  {publisher} {Cambridge University Press},\ \bibinfo
  {year} {2013})\BibitemShut {NoStop}%
\bibitem [{\citenamefont {Andrews}\ \emph {et~al.}(2018)\citenamefont
  {Andrews}, \citenamefont {Jones}, \citenamefont {Salam},\ and\ \citenamefont
  {Woolley}}]{Andrews2018}%
  \BibitemOpen
  \bibfield  {author} {\bibinfo {author} {\bibfnamefont {D.~L.}\ \bibnamefont
  {Andrews}}, \bibinfo {author} {\bibfnamefont {G.~A.}\ \bibnamefont {Jones}},
  \bibinfo {author} {\bibfnamefont {A.}~\bibnamefont {Salam}}, \ and\ \bibinfo
  {author} {\bibfnamefont {R.~G.}\ \bibnamefont {Woolley}},\ }\href {\doibase
  10.1063/1.5018399} {\bibfield  {journal} {\bibinfo  {journal} {The Journal of
  Chemical Physics}\ }\textbf {\bibinfo {volume} {148}},\ \bibinfo {pages}
  {040901} (\bibinfo {year} {2018})}\BibitemShut {NoStop}%
\end{thebibliography}

%

\end{document}